\documentclass[twocolumn,showpacs,superscriptaddress,prb]{revtex4}%
\usepackage{amsfonts}
\usepackage{amsmath}
\usepackage{amssymb}
\usepackage{graphicx}%
\begin{document}
\title{Evidence of the virtual Anderson transition in a narrow impurity band of p-GaAs/AlGaAs quantum wells:
$\varepsilon_4$ conductivity and electric breakdown at low temperatures }
\author{N.V.Agrinskaya}
\author{V.I.Kozub}
\author{D.V.Shamshur}
\affiliation{A.F.Ioffe Physico-Technical
Institute, St.-Petersburg 194021, Russia \vspace{0.01in}}

\begin{abstract}
In highly doped uncompensated p-type layers within the central
part of GaAs/AlGaAs quantum wells at low temperatures we observed
an activated behavior of the conductivity with low activation
energies (1-3) meV which can not be ascribed to standard
mechanisms. We attribute this behavior to the delocalization of
hole states near the maximum of the narrow impurity band in the
sense of the Anderson transition. Low temperature conduction
$\varepsilon_4$ is supported by an activation of minority carriers
- electrons (resulting from a weak compensation by back-ground
defects) - from the Fermi level to the band of delocalized states
mentioned above. The corresponding behavior can be specified as
virtual Anderson transition. Low temperature transport ($<4$ K)
exhibits also strong nonlinearity of a breakdown type
characterized in particular by S-shaped I-V curve. The
nonlinearity is observed in unexpectedly low fields ($<10$ V/cm).
Such a behavior can be explained by a simple model implying an
impact ionization of the localized states of the minority carriers
mentioned above to the band of Anderson-delocalized states.

\end{abstract}
 \maketitle

\section{Introduction}

The problem of a conductivity via localized states was addressed
during a long time; however it mostly concerned a hopping
conductivity. At the same time experiment
\cite{1},\cite{2},\cite{3} demonstrated that a transition to
metallic conductivity (or, in the case of 2D structures, to weakly
localized conductivity) can occur within the impurity band still
split from the conductance or valence band. Recent observation of
superconductivity in B-doped diamond which was supported by the
states of impurity band \cite{4} once more stimulated an interest
to the conductivity over localized states within strongly doped
semiconductors.

Note that typically the situation of the compensated samples was
studied where the Fermi level was situated in a region of large
density of states. Metal-insulator transition in this case is
associated with Anderson transition or, at least, with
Mott-Anderson transition. As for the weakly compensated materials,
until now a situation of the metal-insulator transition is not
completely clear. Indeed, without a compensation a finite
conductivity (including a hopping regime) can exist only via the
states of the upper Hubbard band and in this case we deal with a
"pure" Mott transition. However a presence of even small number of
compensating defects allows a finite hopping conductivity over
single-occupied states. Indeed, in this case a few "holes" within
the impurity band, created by the compensating defects, are
activated from the Fermi level to the maximum of the impurity band
(nearest neighbor hopping). At the same time one can expect that
if the concentration of dopants is large enough, then the carrier
states within the impurity band become to be delocalized. Although
the transport of the "majority carriers" is still forbidden due to
strong on-site correlations, the activated "minority carriers"
(arising from the compensating defects) can propagate as it would
be in a metal. So we deal with "virtual" Anderson transition: it
happens in a situation where the delocalized states are still
above the chemical potential, but the delocalized states do not
exist until the concentration does not exceed some critical one.
Since the disorder potential within the weakly compensated samples
is weak enough, it allows to expect that such a "virtual" Anderson
transition can take place for the dopant concentrations
sufficiently lower than for the Anderson transition for the
"majority carriers". It is important that for 2D structures the
compensating defects can be situated outside of the 2D layer which
also decrease the disorder potential.

In the case of n-doped materials the "majority carriers" within
the impurity band are associated with electrons while the
"minority carriers" are associated with empty donors ("holes"). At
the same time for p-doped materials the "majority carriers"
correspond to holes while the "minority carriers" are related to
negative acceptors, that is to electrons. It is important that the
effective mass of the minority carriers is expected to be
positive. Indeed, for the "majority carriers" the impurity band is
nearly completely occupied while the "minority carriers" are
situated close to the top of the corresponding energy band where
the effective mass of the majority carriers is negative.
Correspondingly, for the minority carriers the effective mass has
the opposite sign with respect to the one of the majority
carriers.

The scenario of the metal-insulator transition (of Mott type or of
Anderson type) depends in real systems on the specific parameters
of the material (compensation degree and related scale of the
disorder potential, effective masses of the carriers etc. In
general, the problem now is far from a complete understanding
which makes corresponding studies to be actual.

In our recent publication \cite{5} we reported an observation of
activated conductivity at low temperatures in Be-doped
uncompensated GaAs/AlGaAs single and multiple quantum wells with
unusually small activation energies (more than an order of
magnitude smaller than the Bohr energy of the dopant). It was
shown that this temperature behavior can not be associated with
nearest neighbor hopping ($\varepsilon_3$ conductivity). We
attributed such a behavior to the "virtual" Anderson transition
mentioned above. Namely, we believe that the observed activation
energy corresponds to activation of the "minority carriers" from
the Fermi level to delocalized states near the maximum of the
impurity band. In the present paper we give detailed arguments in
support of our conclusion including our new experimental results
on low temperature low field breakdown effects observed in the
samples demonstrating the behavior mentioned above. To the best of
our knowledge, the breakdown behavior and S-shaped I-V curves
observed in our experiments can not exist in the hopping regime.
At the same time the scale of the electric fields (less than 10
V/cm) does not allow to relate these effects to breakdown to the
valence band which would need much stronger electric fields.

\section{Experiment}

The technique of growth of multilayered structures with a help of
MBE method was described in our  paper \cite{6}. The structures
contained 1,5,20  GaAs quantum wells with widths 15 nm separated
by barriers of ${\rm Al}_{0.3}{\rm Ga}_{0.7}$As with widths  100
nm.  The middle region of the wells was doped by p-type impurities
(Be), the volume impurity concentration was controlled during the
growth and varied from $1\cdot 10^{18}$ atoms/cm$^3$ up to $2\cdot
10^{18}$ atoms/cm$^3$ (Table 1). The critical concentration for a
bulk p-type GaAS is $2\cdot 10^{18}$ cm$^{-3}$, i.e. the
concentrations mentioned above are of the order or some less than
the critical one. The compensation degree $K=N_D/N_A < 0,01$ was
sufficiently small and supposedly controlled by defects situated
at the edges of the quantum wells and within the barriers. All the
samples parameters are given in Table 1.  The column $N_A$ gives
the bulk acceptor concentration which was controlled during
epitaxial doping.

\begin{tabular}{|c|c|c|c|c|c|c|c|}

\hline
  % after \\: \hline or \cline{col1-col2} \cline{col3-col4} ...
  N & number of wells & well width, nm & $p_{300K}, cm^{-2}$ & $N_A ,cm^{-3} $ & $\varepsilon_1$,meV
   & $\varepsilon_4$,meV & $\sigma_0, e^2/h$ \\
  \hline
  581 & 1 & 15 & $1\cdot10^{12}$ & $2\cdot10^{18}$ & 15 & 2 & 0,1\\
  945 & 1 & 15 & $1,5\cdot10^{12}$ & $1\cdot10^{18}$ & 14 & 2,5 & 0,03\\
  946 & 1 & 15 & $1,7\cdot10^{12}$ & $1,2\cdot10^{18}$ & 13 & 1,5 & 0,1 \\
  484 & 5 & 15 & $1,3\cdot10^{12}$ & $1\cdot10^{18}$ & 26 & 3 & 0,02 \\
  485 & 5 & 15 & $1,5\cdot10^{12}$ & $1,3\cdot10^{18}$ & 13, 26 & 2,5 & 0,03\\
  200N1 & 20 & 15 & $2\cdot10^{12}$ & $1,2\cdot10^{18}$ & 25 & 2 & -\\
  200N2 & 20 & 15 & $3\cdot10^{12}$ & $2\cdot10^{18}$ & 16 & lnT & -\\
    \hline
\end{tabular}

The conductivity of  heavily doped sample (200N2) at low
temperatures weakly decreases with a temperature decrease which
corresponds to weak localization regime and has been reported in
\cite{3}. Note that at high temperatures the samples demonstrated
activated behavior of the conductivity and Hall effect with an
activation energy $\varepsilon_1 \sim 16-20$ meV (intermediate
between $\varepsilon_0$ and $\varepsilon_0/2$, where
$\varepsilon_0 \simeq 28 meV$ - is an energy of isolated acceptor
in  GaAs). Note that in the samples suffering Mott transition,
$\varepsilon_F$ is situated between the centers of the upper and
lower Hubbard bands: $\varepsilon_F \simeq \varepsilon_0- U/2$,
where $U$ - is the Hubbard energy which for 2D is lower than
$\varepsilon_0$. Earlier we estimated the binding energy of doubly
occupied state for the wells with a width  15 nm as 10 meV
\cite{6}, i.e. $U=18$ meV; it gives $\varepsilon_F \simeq 19$ meV,
which agrees with observed values of $\varepsilon_1$ .

For the rest of the samples with concentration of acceptors close
but still less than a critical one we observed a pronounced
activated behavior of the conductivity at temperatures (10-1,3K)
characterized by a low activation energy  $\varepsilon_4 \sim 1-3$
meV (Fig.1).

\begin{figure}[h]
\centerline{ \includegraphics[width=6cm]{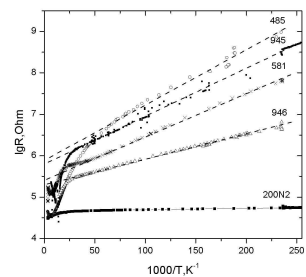}
}
 \caption{Temperature dependences of resistivity for samples with a constant activation energy
 and for sample no. 200N2 that is in the weak localization mode [3].
  \label{fig1} }
\end{figure}

It was most pronounced for the single well structures. The value
of $\varepsilon_4$ decreased with an increase of acceptor
concentration (samples 945,946  ). As it is seen from Fig. 1 and
from the Table, the value of the preexponential $\sigma_0$ was
only by a factor 10 - 30 lower than the universal quantum limit
for 2 D : $\sigma_0 < e^2/h= 4\cdot 10^{-5} Ohm^{-1}$.  Fig. 2
demonstrates temperature behavior of the Hall mobility for these
samples.\vskip6cm

\begin{figure}[h]
\centerline{ \includegraphics[width=6cm]{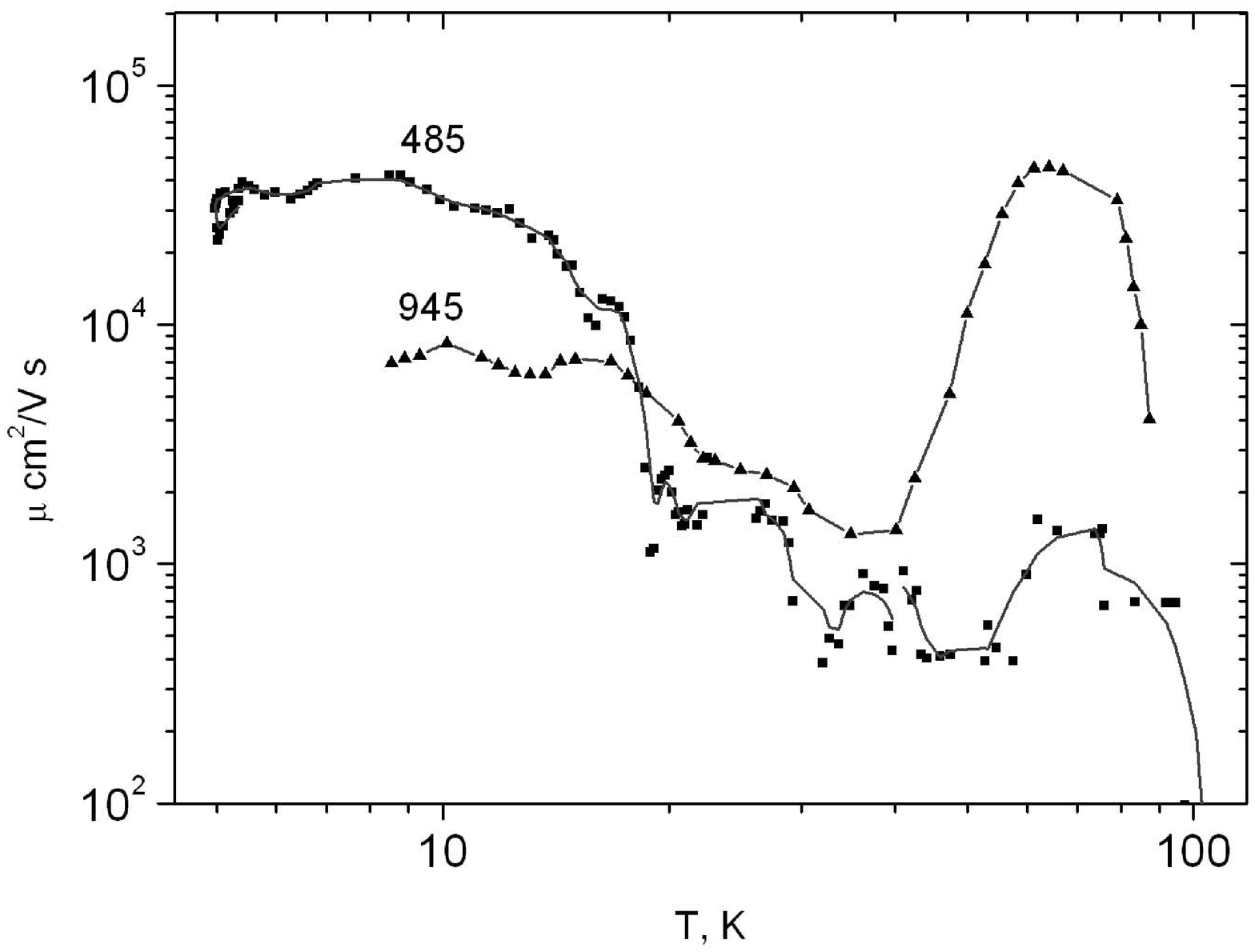} }
 \caption{Temperature dependences of the Hall mobility for several samples with $\varepsilon_4$
  conduction.
  \label{fig2} }
\end{figure}

The maximum observed at high temperatures is attributed to a
competition between scattering by optical phonons ($\mu \propto
T^{-1}$) and scattering by ionized impurities ($\mu \propto T$. At
temperatures 40-10 K one observes a further increase of $\mu$ with
a temperature decrease. Such a behavior is not expected for the
nearest neighbor hopping ($\varepsilon_3$) since in the latter
case the corresponding "mobility" exponentially decreases with a
temperature decrease \cite{7}.

Now let us consider temperature behavior of the conductivity and
Hall effect for these samples at high temperatures 20-300Ê. At $T
= 300 - 50Ê$ there exist an activated behavior resulting from an
activation of the holes from occupied impurity band to the valence
band. The value of $\varepsilon_1$ it is natural to ascribe to a
distance from $\varepsilon_F$ to the percolation level within the
valence band. Since the degree of compensation is small, the width
of the impurity band is also small and, as it will be shown below
$\varepsilon_F - \varepsilon_0 << \varepsilon_0$. Thus at small
enough temperatures $\varepsilon_1 \simeq \varepsilon_0$. However
with the temperature increase when the value of the holes
concentration $p$ starts to be of the order of the dopant
concentration one expects a transition from the regime
$\varepsilon_1 \sim \varepsilon_0$ to the regime $\varepsilon_1
\sim \varepsilon_0/2=14$ meV. It is this behavior which is
observed experimentally  (Table 1). This fact evidences that the
low temperature behavior can not be attributed to $\varepsilon_2$
conductivity.

An important feature is related to the sign of the Hall
coefficient. As it is clearly demonstrated (Fig. 3 ) the sign at
300 K is {\it opposite} to the sign at small temperatures.

\begin{figure}[h]
\centerline{ \includegraphics[width=6cm]{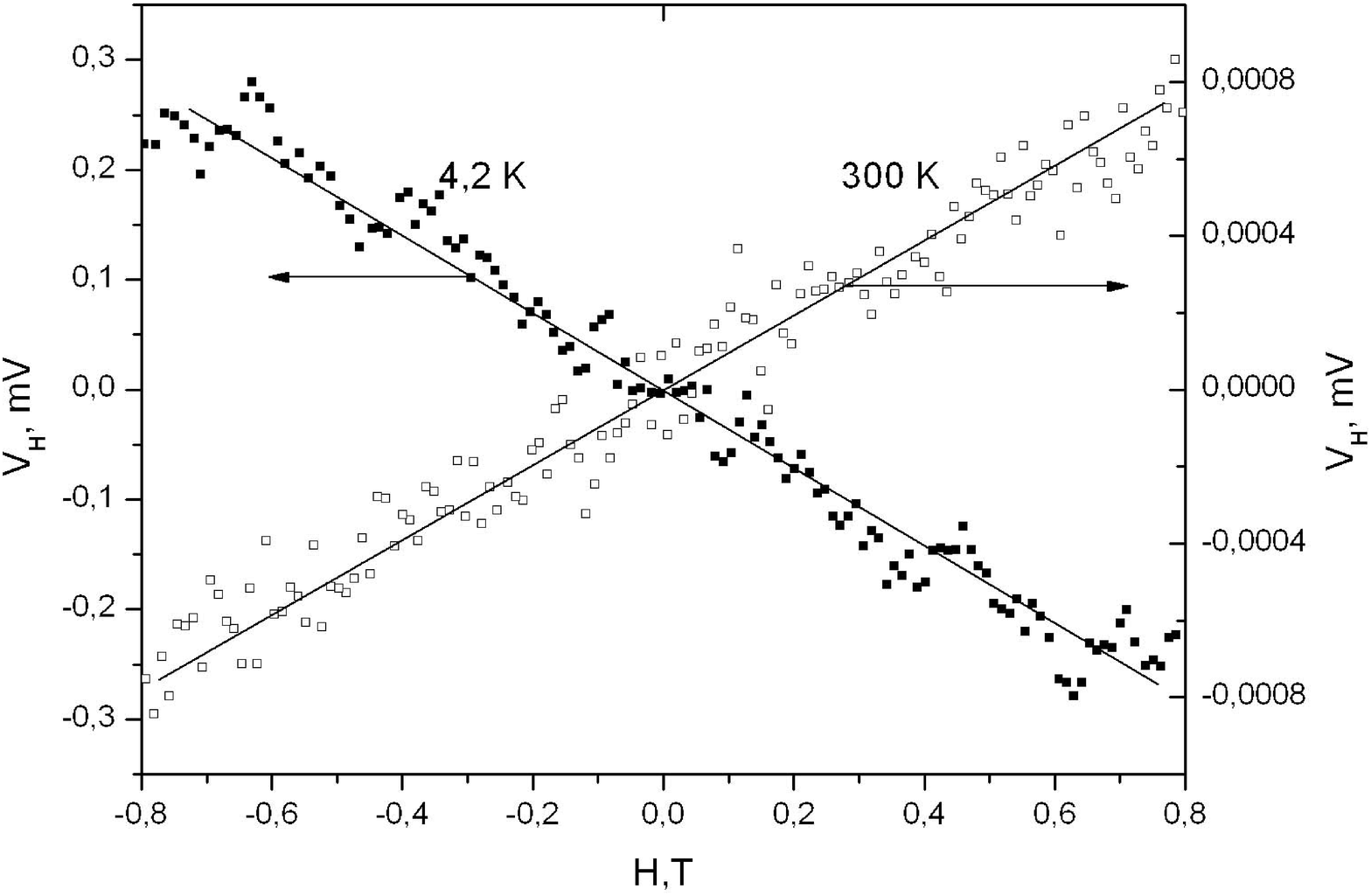} }
 \caption{Hall voltage as a function of magnetic field for $T =
 4.2 K$ and $T = 300 K$ for a fixed direction of the current $I =
 2$ nA
  \label{fig3} }
\end{figure}

At the same time the concentration estimated from the Hall
coefficient appears to be 2-3 orders of magnitude lower than at $T
= 300 $K.

A specific attention was paid to the nonlinear effects in
conductivity. Fig. 4 shows temperature behavior of resistance for
different samples measured in a regime of constant current (0,1 -
2 nA).

\begin{figure}[h]
\centerline{ \includegraphics[width=6cm]{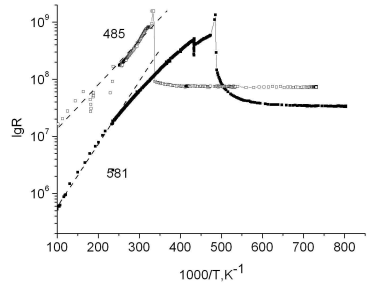} }
 \caption{ Temperature dependences of resistance for different
 samples measured in a regime of constant current (0,1 - 1 nA).
  \label{fig4} }
\end{figure}

It is seen that at low temperatures the samples exhibit a sharp
transition from insulating behavior (Arrenius law)  to a metallic
one; a magnitude of the resistance is decreased by 1.5-2 orders of
magnitude. We believe that transition results from electric
breakdown which takes place when the voltage drop on the sample
(increasing with a temperature decrease) exceeds some critical
value. The temperature of the transition differed for different
samples and was equal to 3Ê, 2,4Ê, 2,1Ê. This parameter is
supposedly controlled by the voltage drop at the given temperature
which depends on the slop of temperature behavior of the
conductivity.

Fig. 5 shows I-V curves of the samples (obtained in the regime of
constant $I$) at different temperatures. The curves clearly
demonstrate a presence of S-shaped regions. One notes that the
breakdown takes place at relatively weak electric fields which can
be estimated as less than 10 V/cm.

\begin{figure}[h]
\centerline{ \includegraphics[width=6cm]{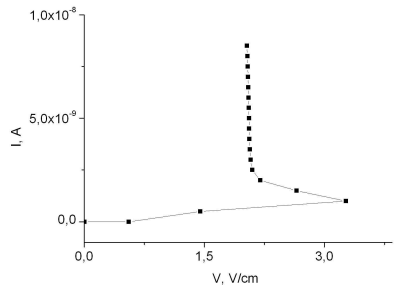} }
 \caption{ I-V curve for sample 485 at 4,2 K.
  \label{fig5} }
\end{figure}

This breakdown can hardly take place in the regime of hopping
conductivity. The hopping conductivity in finite electric fields
was  studied earlier for the similar group of samples \cite{6} .
As it has been shown, the hopping conductivity strongly
(exponentially) increases with an electric filed increase, however
this increase is still a gradual one and no breakdown-like
behavior is observed (which agrees with theoretical predictions -
see e.g. \cite{Shklovskii})

The breakdown can naturally result from an impact ionization of
the acceptors  to the valence band. However in this case the
conductivity is expected to increase much larger than is observed
experimentally. Then, for our case of relatively deep localized
states, the breakdown would be expected at the fields at least
orders of magnitude stronger than is observed experimentally (see
e.g. \cite{Vorobev}.

\section{Discussion}.

As it was noted earlier, the observed Hall mobility increases with
decreasing temperature which excludes $\varepsilon_3$ hopping
conductivity as a mechanism leading to activated behavior of
conductance at low temperatures and leads us to a conclusion that
the activated minority carriers are delocalized. In this case the
increase in the mobility with decreasing temperature can be
explained by the scattering of delocalized carriers by acoustic
phonons. This assumption does not contradict the fact that the
mobility increases with temperature at high temperatures. We
attribute this increase to the contribution of carriers activated
to the valence band, where scattering on charged impurities
prevails at corresponding temperatures. As known, scattering on
acoustic phonons is very sensitive to the effective mass of the
carriers. On the other hand, the effective mass of delocalized
carriers in the impurity band can noticeably exceed the carrier
mass in the valence band as, in particular, was shown in our paper
[3], where this excess was estimated by a factor of 2-3. For this
reason, the changes in the scattering mechanism and a character of
temperature dependence of mobility can be expected with decreasing
temperature.

There is another important evidence to support our picture which
is related to the fact that a sign of Hall effect, observed at 4.2
K, is opposite to the sign of the Hall effect at 300K. The change
of the sign apparently takes place at temperatures 20 - 40 K while
the Hall data at these temperatures are not completely reliable.

 Indeed, with an assumption of an existence of a band of delocalized
 states(BDS)
 within the narrow impurity band we conclude that there are 2 mobility
 edges separating the delocalized states from localized tail
 states. At low temperatures the conducting electrons are mainly
 concentrated near the bottom of the BDS where the effective mass
 of the electrons is expected to be positive since the
 corresponding density of states decreases with energy decrease.

Thus the Hall effect at high temperatures is produced
 by holes with positive effective mass while the Hall effect at
 low temperatures - by electrons with positive effective mass
 which explains the observed behavior.

 Note that the effective mass of an electron near the opposite
 mobility edge (i.e. near the top of BDS) is expected to be
 negative. One has in mind that at high enough temperatures 20-50
 K (which are larger than the activation energy $\varepsilon_4$)
 all the electrons are expected to be distributed over the BDS
 with nearly equal probability and thus the net Hall voltage is
 expected to be small due to different signs of the effective mass
 at different parts of the BDS. It gives an alternative
 explanation of the behavior of Hall mobility which nearly
 vanishes at 20 -40 K but increases with temperature decrease when
 the contribution of electrons with positive effective mass start
 to dominate.

An important evidence supporting our scenario is given by the
character of non-Ohmic behavior. The fact that the conductance in
low-Ohmic state is still much less than high temperature
conductance over the valence band excludes a breakdown to the
valence band. Then, the electric fields in our experiments are
orders of magnitude less than necessary for impact ionization of
deep centers \cite{Vorobev}.

In addition, extremely small electric powers (note that the
breakdown behavior is observed at currents less than 1 nA!) do not
allow any heating effects.

Thus we believe that the behavior observed results from the impact
ionization of the electrons localized in the tail of the impurity
band to the region of delocalized states. The important factor is
a presence of a gap between the Fermi level at and the "mobility
edge" within the impurity band separating strongly and weakly
localized states. This gap is filled by "intermediate" localized
states (see fig. 6) with a concentration much larger than the
electron concentration.

\begin{figure}[h]
\centerline{ \includegraphics[width=6cm]{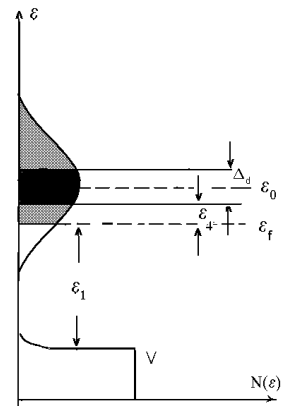} }
 \caption{Scheme of the density of states in the samples with narrow impurity band,
 where $\varepsilon_1$ and $\varepsilon_4$ are the activation energies, $\varepsilon_0$ is
 the  isolated acceptor position, $\varepsilon_f$ is the Fermi energy, and $\Delta_d$ is the
 width of the band of delocalized states.
  \label{fig6} }
\end{figure}

The electrons existing within the BDS (responsible for the impact
ionization) with much larger probability can be trapped by these
"intermediate" states than by their initial centers. Since
electron transport between the localized states is a hopping one,
the return of the electrons to their initial position within the
tail is suppressed. In its turn, the energy positions of these
intermediate states are closer to the mobility edge than the tail
states. As the result, the threshold field for the impact
ionization from these intermediate states is lower that the
threshold field for the ionization of the tails states.
Correspondingly, the low-Ohmic state characterized by a presence
of a finite electron concentration within the BDS can be supported
at the electric fields which are lower than necessary for the
ionization from the deep tail states which explains instabilities
and S-shaped IV curves. (Note that the mechanism of a formation of
S-shaped IV-curves related to a presence of "intermediate" states
between the ionized level and the conductance band was considered
in \cite{Budsulak})

The detailed theoretical treatment supporting the scenario given
above is given in the Appendix.  We have taken into account both
kinetics of impact ionization and nonequilibrium electron
distribution in strong enough fields. The first critical field,
according to our calculations, is given by the estimate
\begin{equation}\label{E1111}
eE_{1} (l_{i} l_{e-ph)})^{1/2} \simeq
\frac{\varepsilon_4}{5^{1/2}}
\end{equation}
Here $l_i = v\tau_i$ is an electron mean free path ($\tau_i$ is
transport relaxation time) while $l_{e-ph} = v \tau_{e-ph}$ where
$\tau_{e-ph}$ is an electron-phonon relaxation time (note that
both $l_i$ and $l_{e-ph}$ correspond to electron energies equal to
$\varepsilon_4$). This equation has a clear physical explanation.
The length $(l_i l_{e-ph})^{1/2}$ is a length of electron
diffusion during its inelastic relaxation time (controlled by
phonons). Thus the critical field $E_1$ allows the energy gain
from the electric field of the order of the energy
$\varepsilon_4$, the latter is just the energy necessary to
activate an electron from the Fermi level to the delocalized
states.

The critical field $E_2$ is given by the similar equation, however
the energy $\varepsilon_4$ is in this case replaced by smaller
energy $\varepsilon_g$ corresponding to a typical distance from
the electron trap states to the mobility edge. Thus we clearly
have $E_1 > E_2$.

Now let us give a rough estimates of the parameters allowing the
scenario mentioned above. Our sample exhibits  an increase of
mobility with temperature decrease nearly saturated at $T \sim
10-15 $ K at the values of the order of $10^4 V^{-1}cm^2 s^{-1}$.
The saturation demonstrates a change of scattering mechanism
(supposedly from electron-phonon scattering to electron-defect
scattering).  Implying a value of $m \sim 10^{-27} $g (which is
several times larger than for the valence band) one estimates the
transport relaxation time as $\sim
 10^{-11}$ s. To estimate $\tau_{e-ph,s}$ one concludes that
 transport relaxation rate $\tau_{i}^{-1}$ at the crossover
 temperature
 only by a factor $1/2$ is contributed by phonons. If we assume that characteristic electron energy is some less than
 this saturation temperature like $ \sim 1$ meV, one concludes
 that $\tau_{e-ph,s} \sim 4 \cdot 10^{-11}$ s since $\tau_{e-ph,s}^{-1} \propto \varepsilon^2$.
Then, for the energy $\sim \varepsilon_m \sim 1$meV one estimates
the electron velocity as $\sim 3\cdot 10^{6}$cm/s . Thus the
critical field $E_1$ can be estimated as of the order of $\sim 10
V/cm$. The electric field $E_2$ is expected to be several times
smaller, first, due to the fact that $\varepsilon_g <
\varepsilon_m$, then also due to a possible role of
electron-electron scattering which can be effective at large
enough concentration of the mobile carriers $n$ (see Appendix).

As it is seen from Fig. 5 , our rough estimates of the threshold
field are in order of magnitude agreement with experimental
results.

Then, from Fig. 3 it is seen that the conductance of the low-Ohmic
state is of the order of the Ohmic conductance at $T \sim 10$ K
and is somewhat smaller than the conductance at intermediate
temperatures 20 - 40 K (when the electrons are effectively
activated to the BDS).  This fact evidences, first, that we do not
deal with impact ionization of the holes to the valence band (the
resulting conductance would be of the order of the high
temperature one which is several order of magnitude higher). Then,
it shows that only a fraction of localized electrons is ionized to
the BDS.

Note that the estimated threshold field is surprisingly low. We
know only few references reporting S-shaped I-V curves at such a
low fields, but for the case of shallow impurities [ ]. Since in
our case we deal with relatively deep acceptors we can conclude
that the observation of S-shaped I-V curves at low fields
evidences a vicinity of metal-insulator transition which can not
be of the Mott type. Indeed, it would contradict to the
high-temperature activation energies which almost coincide with
ionization energies of isolated acceptor while in a situation of
the Mott transition Fermi level would be situated between energies
of single occupied and doubly occupied states. In combination with
the previous data it gives a strong support to the scenario of
virtual Anderson transition. Then, it also allows to conclude that
the observed $\varepsilon_4$ conductivity can not be attributed to
$\varepsilon_3$ channel.

Thus, the results indicate that activation occurs to the band of
the delocalized states appearing due to the Anderson transition in
the impurity band. It is worth noting that the Anderson transition
is typically considered in the single-particle picture in the
absence of electron-electron correlations. In our case, the
repulsion of holes from acceptors prevents transport of
delocalized holes  even if the wave functions in single-particle
approximation are delocalized. However, owing to the activation of
electrons from the Fermi level located at the tail of the impurity
band to the indicated states, transport through such delocalized
states becomes possible. In contrast to the real metal-insulator
transition (when the Fermi level is located in the
delocalized-state region), the manifestation of the delocalized
states in our case can be described as the virtual Anderson
transition. As we have shown earlier \cite{5}, in the case of
narrow impurity band existing in non-compensated samples due to
the weak disorder potential the critical concentration of such a
virtual Anderson transition is smaller than given by the Mott
criterion by the factor
\begin{equation}
\ln^{-2} (\varepsilon_0/\Delta \varepsilon)
\end{equation}
where $\Delta \varepsilon << \varepsilon_0$ is a width of the
impurity band. Correspondingly, although the dopant concentration
in our samples was less than the critical one for the real
Mott-Anderson transition, it can readily be larger than the
critical concentration for the virtual Anderson transition.

There is a peculiar question concerning localization according to
Lifshits scenario \cite{Lifshits}. Indeed, the scatter in inter
impurities distances can impose localization as well as a scatter
of energies and the corresponding criterion is except a numerical
coefficient is similar to Mott criterion. However, first, the
coefficient for the Lifshits transition is still unknown and we
can expect that the Lifshits localization starts to be ineffective
at lower concentrations than Anderson localization. Then, any
deviation from purely random distribution of impurities (which can
occur in course of doping procedure) suppress Lifshits
localization.

To conclude, we have given experimental and theoretical evidences
of an existence of virtual Anderson transition, characterized by
an appearance of delocalized states within an impurity band above
the chemical potential. Such a transition is expected to be
typical for strongly doped, but uncompensated materials and
precedes (as a function of dopant concentration) standard
Mott-Anderson transition. In our experimental studies of highly
doped uncompensated p-type layers within the central part of
GaAs/AlGaAs quantum wells it manifested itself, in particular, in
activated behavior of low temperature conductivity. We have shown
that the latter can not be associated with known $\varepsilon_2$
or $\varepsilon_3$ mechanisms but results from an activation of
electrons from the chemical potential to the delocalized states
mentioned above. The fact that the type of the carriers in this
case is different from the one supporting conductivity within the
valence band is evidenced by a change of the sign of Hall
coefficient with a temperature decrease. Another evidence of the
virtual Anderson transition is given by electric breakdown
observed at low temperature in unusually weak (for relatively deep
centers) electric fields. We believe that this behavior is related
to impact ionization of the localized electrons from below the
chemical potential to the band of Anderson-delocalized states.

\section{Acknowledgements}
We are grateful to A.E. Zhukov for manufacturing the structures,
to A.S. Ioselevich for discussion. This work was supported by the
Russian Foundation for Basic Research (project no. 06-02-17068).

\section{Appendix}

We consider impact ionization of localized "minority" carriers -
electrons - situated deep in the tail of the impurity band (with a
concentration $N_1$, controlled by compensating defects) by
delocalized minority carriers to delocalized states with energies
higher than the mobility edge $\varepsilon_m$ (separating strongly
and weakly localized states). We also have into account that in
between $\varepsilon_m$ and the position of the chemical potential
$\mu $ at the equilibrium there exists a wide region of localized
states which at low temperatures are occupied by holes and thus
they can trap mobile electrons; the total number of such localized
states $N_2$ is much larger than $N_1$. While the analytical
solution of the problem with continuous spectrum of the localized
states seems to be impossible, we will model our system as the
3-level one, including the "tail" states, the intermediate level
of localized states separated from delocalized states by some
energy $\varepsilon_g < \varepsilon_m$, and the band of
delocalized states.

Denoting the the number of electrons captured on the localized
states as $\tilde n$ we describe the process of impact ionization
at small temperatures as follows:
\begin{eqnarray}\label{ion}
\frac{{\rm d} n}{{\rm d} t} = - B_R n (N_2 + n) + A_I n {\cal
F}_2{\tilde n} + A_I n {\cal F}_1(N_1 - {\tilde n} - n)
\nonumber\\
\frac{{\rm d} {\tilde n}}{{\rm d} t} = - {\tilde B}_R {\tilde n}(n
+ {\tilde n}) -   A_I  {\cal F}_2n{\tilde n} + B_R n(N_2 - {\tilde
n})
\end{eqnarray}
($B_R$ is the coefficient describing a recombination of the
electrons to any localized states, ${\tilde B}_R $ is the
coefficient describing the recombination of electrons from
intermediate states to their initial positions, $A_I$ is a
coefficient of impact ionization, ${\cal F}_1$ and ${\cal F}_2$
are the coefficients describing the relative numbers of mobile
electrons with kinetic energies larger than $\varepsilon_m$ and
$\varepsilon_g$, respectively.

 Note that in principle the coefficients $A_I$,
$B_R$ are energy dependent. However we will neglect such a
dependence since the width of the impurity band is not much larger
than $\varepsilon_m$ and thus the relative change of $\varepsilon$
at energies higher higher than the threshold value $
\varepsilon_m$ is not large.

Analyzing the first of Eqs.\ref{ion} one concludes that since $N_2
>> N_1$ if initially $n = 0$, $\tilde n =
0$, this solution is stable with respect to small fluctuations up
to threshold electric field $E_1$ corresponding to
\begin{equation}\label{E1}
A_I {\cal F}_1 N_1 = B_R N_2
\end{equation}
At the same time for finite $n$, $\tilde n$  the solution of
Eq.\ref{ion} gives
\begin{eqnarray}\label{2}
{\tilde n} \simeq \frac{B_R N_2}{A_I {\cal F}_2}\nonumber\\
n = N_1 - {\tilde n}
\end{eqnarray}
Here we make use of the relations $A_I {\cal F}_2 >> B_R$ (which
holds at electric fields of the order of $E_1$ ), $\tilde B_R <<
B_R$, $N_2
>> N_1$.
This solution is stable until ${\tilde n} \leq N_1$ which holds
until
\begin{equation}\label{E2}
A_I{\cal F}_2N_1 \geq B_R N_2
\end{equation}
which defines another critical field $E_2$. Since $ \varepsilon_m
> \varepsilon_g$ it is expected that for a given field ${\cal F}_2 >> {\cal
F}_1$. Thus the value of $E_2$ can me smaller than $E_1$
 and thus we deal with two branches of the solution: the
one corresponding to $n =0$, $\tilde n = 0$ which is stable with
respect to small fluctuations at $E < E_1$ and the solution given
by Eq.\ref{2}. For electric fields $E \sim E_1$ it corresponds to
practically complete ionization of the electrons to the
delocalized states. However this solution can exist at smaller
fields until they are larger than $E_2$, that is  $E_2 < E < E_1$
the two solutions coexist and are stable with respect to small
fluctuations. We will give more detailed analysis later with an
account of the fact that both the function ${\cal F}$ and
coefficients $B_R$ can be different for the two branches.

Now let us calculate the principal functions ${\cal F}_1$ and
${\cal F}_2$. Considering the delocalized electrons we will take
the mobility edge as the origin for $\varepsilon$. By definition
\begin{equation}
{\cal F}_1 = n^{-1} \int_{\varepsilon_m} {\rm d} \varepsilon \nu
(\varepsilon) f(\varepsilon), \hskip1cm {\cal F}_2 = n^{-1}
\int_{\varepsilon_g} {\rm d} \varepsilon \nu (\varepsilon) f
(\varepsilon)
\end{equation}
where $f$ as an electron distribution function.

We would like to emphasize that, strictly speaking, the electron
transport in our situation can hardly be described in a same way
as for standard energy band - at least in the vicinity of the
mobility edge, so our simple equations can be considered as
semiqualitative. For relatively small $n$ (when electron-electron
processes are not effective) we will use a concept of energy
diffusion and write with a neglect of electron-electron scattering
\begin{equation}\label{gen}
(D_E + \frac{(\hbar
\omega_T)^2}{\tau_{e-ph,T}})\nabla_{\varepsilon} f + \frac{(\hbar
\omega_s)}{\tau_{e-ph,s} }f = 0
\end{equation}
Here the first term describes a diffusion of an electron along
energy axis due to interactions with thermal phonons with energies
$\hbar \omega_T = T$ (phonon absorption and stimulated emission
characterized by the relaxation time $\tau_{e-ph,T}$) and due to
energy gains and energy losses in course of chaotic motion in
electric field,
\begin{equation}
D_E \sim \frac{(eEv\tau_i)^2}{\tau_i} \hskip1cm,
\end{equation}
$\tau_i$ is momentum relaxation time. The second term describes an
energy drift along energy axis due to spontaneous emissions of
phonons with typical frequency $\hbar \omega_s$ characterized by
relaxation time $\tau_{e-ph,s}$. The electron-phonon relaxation
for 2D electrons it is affected by the fact that the normal
component of the phonon momentum is not conserved \cite{Karpus}.
According to \cite{Karpus} at low temperatures $\tau_{e-ph,T}^{-1}
\propto T^3 \varepsilon^{-1/2}$ which is similar to
electron-phonon relaxation time in 3D metals and $\hbar \omega_T
\sim T$. At the same time for relatively small energies
$\varepsilon < (ms^2 W)^{1/2}$ (where $s$ is a sound velocity
while $W$ is an energy of lateral quantization which in our case
of 2D impurity band is of the order of the Bohr energy)
$\tau_{e-ph,s}^{-1} \propto \varepsilon^2$ while $\hbar \omega_s
\sim \varepsilon$.

The solution of Eq.\ref{gen} depends on the relation between $D_E$
and phonon contribution to the energy diffusion. If the last term
dominates, the solution corresponds to equilibrium, $f \propto
\exp - (\varepsilon/T)$. If we have the opposite relation, which
holds when
\begin{equation}\label{non}
(eE)^2 \frac{v^2 \tau_i \tau_{e-ph,T}}{T^2}
>  1
\end{equation}
we deal with strongly nonequilibrium distribution
\begin{equation}
f \propto \exp - \int_0^{\varepsilon} {\rm d} \varepsilon'
\frac{\varepsilon}{(eE)^2 v^2 \tau_i \tau_{e-ph,s}}
\end{equation}
According to our experimental data on temperature dependence of
the mobility, at the energy region of interest $\varepsilon \sim
\varepsilon_m$ $\tau_i \propto \varepsilon^{-d}$ where $d \simeq
2$ and thus
\begin{equation}\label{nonexp}
f \propto \exp - \frac{\varepsilon^2}{5 (eE)^2 v^2 \tau_i
(\varepsilon)\tau_s(\varepsilon)}
\end{equation}
This function is nearly constant at $\varepsilon < \varepsilon_1$
where
\begin{equation}\label{varepsilon1}
\frac{5(eE)^2 v^2 \tau_i
\tau_{e-ph,s}}{\varepsilon^2}|_{\varepsilon_1} = 1
\end{equation}
and extremely steeply (much stronger than simple exponential)
decays at $\varepsilon > \varepsilon_1$. As it is seen, for strong
enough fields when $\varepsilon_1 >> T$ the inequality of
Eq.\ref{non} holds automatically.

For the further analysis we should have estimates for $B_R$ and
$A_I$. To estimate $A_I$ we have in mind that $A_I = v\sigma_I$
where $\sigma_I$ is the ionization cross-section. The latter (in
2D) can be very roughly estimated as $\sigma_C
(\sigma_t^2/\lambda^2)$ where $\sigma_C \sim (e^2/(\kappa
\varepsilon))^{1/2} $  is the geometrical cross-section of a
Coulomb scattering of an electron with kinetic energy $\varepsilon
> \varepsilon_m$ by a charged center, $\kappa$ being a dielectric
constant, $\sigma_t$ is a cross-section of an elastic scattering
of an electron by the trap center and $\lambda$ is a typical
electron wavelength which can be roughly estimated as
$\lambda^{-2} \sim N_3$ where $N_3$ is a concentration of the
delocalized states. Note that $\sigma_t$ can be estimated from
above  basing on the known values of electron mean free path $l_i$
as $\sigma_t \sim (N_2l_i)^{-1}$ (actually the corresponding
cross-section can be smaller if electron momentum relaxation is
dominated by other mechanism). As for $B_R$ we shall consider
recombination due to phonon emission. In this case recombination
process involves simultaneous interaction of an electron with
phonon and trap center (neutral). One can estimate the
recombination cross-section $\sigma_R \sim B_R/v$ as
$\sigma_t^2/v\tau_{e-ph,s}$.

Correspondingly, the estimates of the critical fields $E_1$
($E_2$) can be rewritten as
\begin{equation}\label{E11}
\frac{B_R N_2}{A_I {\cal F}_{1(2)}} N_1 \simeq \frac{N_2}{N_3N_1
v\tau_{e-ph,s} \sigma_C {\cal F}_{1(2)}} = 1
\end{equation}
Assuming that $N_2 \sim N_3$, one concludes that our scenario can
hold provided
\begin{equation}\label{cri}
N_1 v\tau_{e-ph,s} \sigma_C > 1
\end{equation}
since in any case ${\cal F} < 1 $. As it can be readily estimated,
$\sigma_C \sim 2 \cdot 10^{-5}$cm while $v \tau_{e-ph,s} \sim
10^{-4}$ cm. Thus this criterion is obeyed for $N_1 > 0.5 \cdot
10^{9}$ cm$^-2$.

Since an equilibrium phase at $T \rightarrow 0$ corresponds to $n
= 0$, the threshold field and, correspondingly, ${\cal F}_1$, is
related to a distribution given by Eq.\ref{nonexp} which can be
modelled as a cut-off of  of $f(\varepsilon)$ at $\varepsilon =
\varepsilon_1$. Correspondingly, ${\cal F}_1 \simeq \theta
(\varepsilon_1 - \varepsilon_m) (\varepsilon_1 -
\varepsilon_m)/\varepsilon_1$. Thus $E_1$ can be estimated from an
equality $\varepsilon_1 \simeq \varepsilon_m$ which gives
\begin{equation}\label{E111}
eE_1 (l_{i,m} l_{e-ph,m})^{1/2} \simeq
\frac{\varepsilon_m}{5^{1/2}}
\end{equation}
where $l_{i,m} = (v \tau_i)|_{\varepsilon = \varepsilon_m}$,
$l_{e-ph,m} = (v \tau_{e-ph,s})|_{\varepsilon = \varepsilon_m}$.
Indeed, for $E > E_1$ ${\cal F}_1 \sim 1$ and the condition of
Eq.\ref{E11} holds. Note that $\varepsilon_m$ by definition
coincides with the activation energy $\varepsilon_4$.

As it is seen, the estimate of $E_2$ is a similar one and we have
$E_2 < E_1$  because of the replacement of $\varepsilon_m$ by
$\varepsilon_g < \varepsilon_m$ in Eq.\ref{E111}.

If $n$ is high enough, the energy relaxation is dominated by
electron-electron rather than by electron-phonon processes and
thus
\begin{equation}\label{Te}
f = \exp - \varepsilon/T_e
\end{equation}
However the energy transfer from the electron system to the bath
is still due to electron-phonon processes. The effective electron
temperature $T_e$ which is established in the system can be
derived from the energy balance between energy gain from electric
field $\sigma E^2 = n(eE)^2 \tau_i/m$  and its decay to the
thermal bath $\sim n (T_e-T) /\tau_{e-ph}$. Thus we have
\begin{equation}
E^2 \gamma (T_e) = \frac{T_e}{T} - 1
\end{equation}
As it is seen from Eq.\ref{Te} and Eq.\ref{varepsilon1}, formally
the estimate for $T_e$ except of a numerical factor coincides with
estimate for $\varepsilon_1$ and characterizes an average electron
energy which is controlled by energy balance between electrons and
the lattice and does not depend on the efficiency of
electron-electron processes. However the asymptotic behavior at
large energies ($> T_e, \varepsilon_1$ ) strongly depend on these
processes. Namely, without these processes the decay of
$f(\varepsilon)$ is significantly stronger within the region
$\varepsilon_1 < \varepsilon $.

Let us assume that for electron concentrations $n = N_1$
electron-electron processes are much more effective for energy
relaxation than electron-phonon ones. Thus the second branch
(where $n \sim N_1$) corresponds to effective electron-electron
scattering and electron distribution is described by Eq.\ref{Te}.
Correspondingly, ${\cal F}_2 = \exp ( - \varepsilon_g/T_e)$. Thus
the electric field $E_2$ is given by the equation
\begin{equation}\label{E22}
T_e(E_2) = \varepsilon_g \ln^{-1} ( \frac{ A_IN_1}{B_RN_2})
\end{equation}

However for the second branch where electron-electron scattering
is assumed to be effective, we shall also consider Auger
processes. Following the same rough procedure as applied above, we
can estimate the cross-section $\sigma_R$ as $\sim (\sigma_t)^2
\sigma_C n$. Note that in this case Eq.\ref{2} has a solution
\begin{equation}
n = \frac{N_1}{1 + N_2/(N_3 {\cal F}_2)}
\end{equation}
and thus Eq. \ref{E2} holds automatically. However one has in mind
that the Auger processes dominate only if
\begin{equation}
v\tau_{e-ph,s} > (n\sigma_C)^{-1}
\end{equation}
which holds only for large enough $n$ and is violated at some
critical $n = n_c$. At $n < n_c$ electron-phonon scattering
dominates and the concept of electron temperature is not valid, so
the system can not stay at the second branch. Thus the estimate of
Eq.\ref{E22} can be rewritten as
\begin{equation}
T_e(E_2) = \varepsilon_g \ln^{-1} ( \frac{N_3N_1}{N_2n_c} )
\end{equation}
Note that the electron-electron processes can lead to additional
source of multistability. Indeed, the concentration of high energy
electrons given by Eq.\ref{Te} can be at the same electric field
larger than given by Eq.\ref{nonexp}. Correspondingly, the
effective impact ionization by electron distribution of Eq.
\ref{Te} can be supported at electric field somewhat weaker than
necessary to support the distribution of Eq.\ref{nonexp}

\end{document}